 \documentclass[final,5p,times,twocolumn]{elsarticle}

\usepackage{amssymb}
\usepackage{amsmath}
\usepackage{amsthm}
\usepackage{xcolor, color, framed}
\usepackage{ulem}
\usepackage[colorlinks, linkcolor=red,anchorcolor=green,citecolor=blue]{hyperref}

\hypersetup{
  pdftitle={Lattice QCD study on nucleon-Omegaccc interaction at the physical point},
  pdfauthor={Liang Zhang}
}
\usepackage{tikz,xcolor}
\usepackage{booktabs}

\newcommand{\graphsize}{0.85}

\graphicspath{{./figures/}}

\journal{Journal of Subatomic Particles and Cosmology}

\begin{document}

\begin{frontmatter}



\title{Lattice QCD study on nucleon-\texorpdfstring{$\Omega_{\rm ccc}$}{Omegaccc} interaction at the physical point}


\fntext[coll]{for HAL QCD Collaboration}
\author[first,second]{Liang Zhang\fnref{coll}}
\ead{zhangliang@sinap.ac.cn}

\affiliation[first]{organization={Shanghai Institute of Applied Physics, Chinese Academy of Sciences}, 
            addressline={Shanghai, 201800, China}}
\affiliation[second]{organization={Key Laboratory of Nuclear Physics and Ion-beam Application (MOE), Institute of Modern Physics, Fudan University},
            addressline={Shanghai, 200433, China}}

\begin{abstract}
The S-wave $N-\Omega_{\rm ccc}$ interaction is investigated using (2+1)-flavor lattice QCD at physical point. By applying the time-dependent HAL QCD method, we identify overall attractive potentials in both the $^3\mathrm{S}_1$ and $^5\mathrm{S}_2$ channels. The corresponding scattering parameters are
$a_0 = 0.56(0.13)\left(^{+0.26}_{-0.03}\right)$ fm and $r_{\mathrm{eff}} = 1.60(0.05)\left(^{+0.04}_{-0.12}\right)$ fm for the spin-1 channel, and 
 $a_0 = 0.38(0.12)\left(^{+0.25}_{-0.00}\right)$ fm and $r_{\mathrm{eff}} = 2.04(0.10)\left(^{+0.03}_{-0.22}\right)$ fm for the spin-2 channel,
 indicating the absence of a bound dibaryon.
 To probe the underlying dynamics, we decompose the interaction into a dominant attractive spin-independent part and a short-range attraction (repulsion) spin-dependent part in the spin-1 (spin-2) channel. Finally, we provide qualitative comparisons with $N-J/\psi$ and $N-\Omega_{\rm sss}$ systems studied previously at $m_\pi \simeq 146$ MeV, to clarify the governing interaction mechanisms.
\end{abstract}




\end{frontmatter}




\section{Introduction}
\label{introduction}

According to Quantum Chromodynamics (QCD), hadrons are color-neutral particles due to the color confinement. However, there are non-trivial interactions between color-neutral hadrons, which cannot be derived by single-hadron properties. 
Dihadron systems with distinct valence quark flavor, exhibiting fully attractive interactions in the absence of quark Pauli blocking~\cite{Sekihara:2018tsb,HALQCD:2018qyu,Lyu:2022imf,Lyu:2024ttm},  content offer a unique opportunity to probe long-range spin–isospin independent forces as well as short-range interactions. 
Especially, the $N-\Omega_{\rm sss}$ system is predicted to have a quasi-bound state near unitarity in $^5{\rm S}_2$~\cite{HALQCD:2018qyu,Sekihara:2018tsb}. 
And the system has also been intensively studied via femtoscopic correlation measurements in heavy-ion and proton-proton collisions~\cite{ALICE:2020mfd,kehao_zhang_search_2025}, which provide evidence for an attractive interaction.

Inspired by the studies of $N-\Omega_{\rm sss}$ system, the possibility of forming a bound state in the $N-\Omega_{\rm ccc}$ channel has been explored. A quark-model calculation has suggested a potential bound state in this system~\cite{Huang:2020bmb}. In this context, first-principles calculations can provide a valuable predictions for this triply charmed states. 
Among all dibaryon systems with charm number $C=3$, the $N-\Omega_{\rm ccc}$ constitutes the lowest threshold (approximately 5740 MeV), well below the $\Lambda_{\rm c}-\Xi_{\rm cc}$ threshold (5910 MeV) and the $\Sigma_{\rm c}-\Xi_{\rm cc}$ threshold (6080 MeV). The resulting $\sim200$ MeV energy gap provides an ideal, uncontaminated environment for investigating the low-energy $N\Omega_{\rm ccc}$ interactions.

In this report, we employ the HAL QCD method~\cite{Ishii:2006ec,Aoki:2020bew}, a non-perturbative approach that determines hadron-hadron interactions directly from spacetime correlation functions, to study the $N-\Omega_{\rm ccc}$ interaction in the two spin-channels ($^3{\rm S}_1$ and $^5{\rm S}_2$)~\cite{Zhang:2025zaa}. The analysis is performed using $(2+1)$-flavor lattice QCD with physical light-quark masses ($m_{\pi}\simeq137.1~\text{MeV}$) and the physical charm-quark mass. 
Lattice QCD data enable a exploratory study of the quark-mass dependence of the interaction between $N-\Omega_{\rm ccc}$ and $N-\Omega_{\rm sss}$, as well as a separation of spin-dependent and spin-independent forces in the $^3{\rm S}_1$ and $^5{\rm S}_2$ channels. This separation offers a useful diagnostic of the underlying long-range dynamics, since the soft-gluon exchange mechanism \cite{Dong:2022rwr} is expected to be predominantly spin independent. 
In addition, by comparing the long-distance behaviors of the spin-independent potential for $N-\Omega_{\rm ccc}$ and that for $N-J/\psi$, we provide qualitative insights into the chromo-polarizability of heavy hadrons.

\section{HAL QCD method}
\label{HAL-QCD}

We briefly summarize the time-dependent HAL QCD method~\cite{Ishii:2012ssm} used to determine an energy-independent non-local potential $U(\boldsymbol{r},\boldsymbol{r}')$ from lattice correlation functions:
{\small
\begin{equation}
    \label{eq:HAL}
    \left(-\partial_t +\frac{1+3\delta^2}{8\mu}\partial_t^2-\hat{H}_0\right)R^J(\boldsymbol{r},t)
    =\int{d^3\boldsymbol{r}'U(\boldsymbol{r},\boldsymbol{r}')R^J(\boldsymbol{r}',t)}.
\end{equation}
}
Here $\hat{H}_0\equiv-\nabla^2/(2\mu)$ with the reduced mass $\mu$ of the $N\Omega_{\rm ccc}$ ($\Omega_{\rm 3c}$) system, and $\delta\equiv(m_{\Omega_{\rm 3c}}-m_N)/(m_{\Omega_{\rm 3c}}+m_N)$.

The normalized four-point function $R^J$ is constructed as
\begin{equation}
\label{eq:4pt}
\begin{aligned}
    R^J(\boldsymbol{r},t)
    &=\frac{
         P_{A^+_1}\sum_{\boldsymbol{x}}
            \left\langle0\right| 
                [
                    N(\boldsymbol{x},t)
                    \Omega_{\rm 3c}(\boldsymbol{r}+\boldsymbol{x},t)
                ]P^s
                \bar{\mathcal{J}}_{N\Omega_{\rm 3c}}(0) 
            \left|0\right\rangle
    }
    {
        G_N(t)G_{\Omega_{\rm 3c}}(t)
    }\\
    &=\sum_{n,M} A^{JM}_n\psi_{N\Omega_{\rm 3c}}^{JM,n}(\boldsymbol{r})e^{-\Delta W_n t},
\end{aligned}
\end{equation}
where $\bar{\mathcal{J}}_{N\Omega_{\rm 3c}}(0)$ is a wall-type six-quark source operator with zero orbital angular momentum ($L=0$), and $G_B(t)\propto \exp{(-m_Bt)}$ is the single-baryon correlator. 
The second line represents the spectral decomposition in terms of the Nambu–Bethe–Salpeter (NBS) wave function $\psi_{N\Omega_{\rm 3c}}^{JM,n}(\boldsymbol{r})$ for the $n$-th energy eigenstate with energy $W_n$ and total angular momentum $(J,M)$, and $\Delta W = W_n - m_{\Omega_{\rm 3c}} - m_N$. $N(\boldsymbol{x},t)$ and $\Omega_{\rm 3c}(\boldsymbol{r}+\boldsymbol{x},t)$ denote the nucleon and $\Omega_{\rm 3c}$ baryon operators, respectively. The $A_1^+$ projection enforces the S-wave component, while the spin projection selects the $J=s$ state.

At the leading order of the derivative expansion, $U(\boldsymbol{r},\boldsymbol{r}')\simeq V^{J}_{\rm LO}(r)\,\delta^{(3)}(\boldsymbol{r}-\boldsymbol{r}')$, the effective central potential reads
\begin{equation}
    \label{eq:local-pot}
    \begin{aligned}
        V^{J}_{\rm LO}(r)
            &=V_0(r)+\boldsymbol{S}_{N}\cdot\boldsymbol{S}_{{\Omega_{\rm 3c}}}V_{s}(r)\\
            &\simeq R^J(\boldsymbol{r},t)^{-1}\left(-\partial_t +\frac{1+3\delta^2}{8\mu}\partial_t^2-\hat{H}_0\right)R^J(\boldsymbol{r},t).
    \end{aligned}
\end{equation}
Here $V_0$ and $V_s$ denote the spin-independent and spin-dependent potentials, respectively. 

\section{Lattice setup}
\label{lattice-setup}

We use the ``HAL-conf-2023'' gauge configuration set (``F-conf'')~\cite{Aoyama:2024cko}, generated in $(2+1)$-flavor lattice QCD on the Fugaku supercomputer at RIKEN. The simulations employ the Iwasaki gauge action at $\beta=1.82$ and the nonperturbatively $O(a)$-improved Wilson quark action with stout smearing, on a $96^4$ lattice at physical quark masses ($m_{\pi}\simeq137~\text{MeV}$). The lattice spacing is $a\simeq0.084$~fm, corresponding to a spatial extent of $L\simeq8.1$ fm.

Charm quarks are treated using the relativistic heavy quark (RHQ) action~\cite{Aoki:2001ra}. Two parameter sets, Set1 and Set2, from Ref.~\cite{Namekawa:2017nfw} are employed to interpolate to the physical charm quark mass, reproducing the dispersion relation of the spin-averaged 1S charmonium state.

For the $N\Omega_{\rm 3c}$ system, quark propagators are solved with a wall-type source with Coulomb gauge fixing, together with periodic boundary conditions in all directions. The analysis is based on 1,600 gauge configurations, with statistical enhancement achieved by averaging over forward and backward propagations, lattice rotations, and multiple source locations, resulting in $4.1\times 10^5$ measurements per parameter set. Statistical uncertainties are estimated using the jackknife method.

To suppress contamination from higher partial waves after the $A^+_1$ projection, we apply Misner’s method~\cite{Misner:1999ab} for approximate partial-wave decomposition on the cubic lattice, which efficiently isolates the S-wave component~\cite{Miyamoto:2019jjc}.

\section{Numerical results}
\label{results}

\subsection{\texorpdfstring{$N-\Omega_{\rm 3c}$}{N-Omega3c} potential in different spin channels}
\label{sub:pot}

\begin{figure}[t]
    \centering
    \includegraphics[width=\graphsize\linewidth]{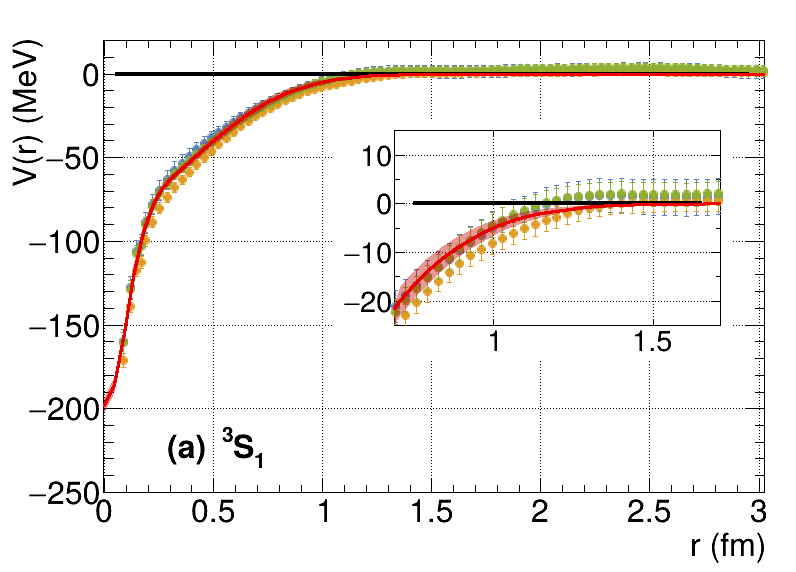}


    \includegraphics[width=\graphsize\linewidth]{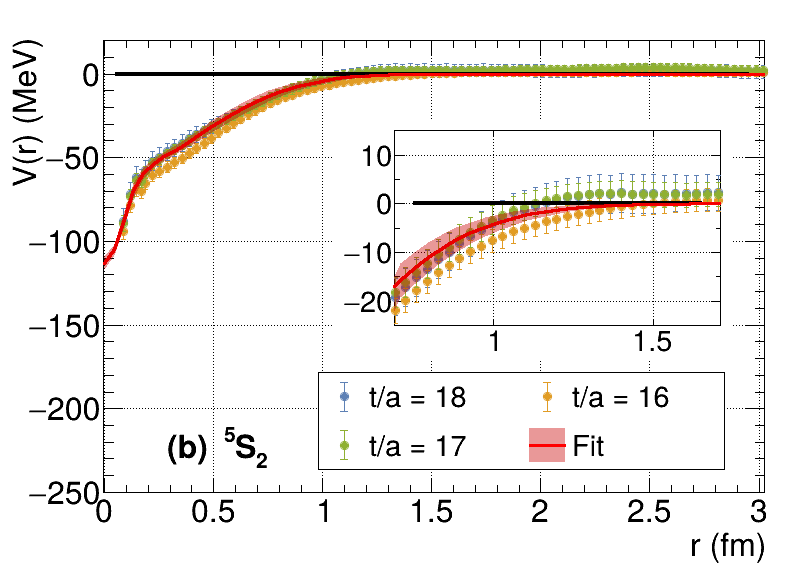}
    \caption{$N-\Omega_{\rm ccc}$ potential in the $^3\mathrm{S}_1$ (a) and $^5\mathrm{S}_2$ (b) channels. The potential with staistical errors is extraced from lattice data at $t/a=16$ (yellow), $17$ (green) and $18$ (blue). Two-range Gaussian $V_{\rm fit}$ fits (purple) at $t/a=17$ in the range of $0<r<3$~fm are drawn in both panels. The figure is adapted from our published work, Ref.~\cite{Zhang:2025zaa}.}
    \label{fig:pot-spin-channels}
\end{figure}

Fig.~\ref{fig:pot-spin-channels} shows the $N-\Omega_{\rm 3c}$ potentials in $^3{\rm S}_1$ and $^5{\rm S}_2$ channels at different time slices $t/a=16,~17,~\text{and}~18$. The results are obtained by interpolating the potentials from Set1 and Set2 to the physical charm quark mass. In both spin channels, the potentials are attractive over the entire distance range and show only a mild time-slice dependence. This qualitative behavior closely resembles those observed in the $N-\Omega_{\rm 3s}$~\cite{HALQCD:2018qyu}, $N-\phi$~\cite{Lyu:2022imf} and $N-{\rm\bar{c}c}$~\cite{Lyu:2024ttm} systems. Such universal attraction can be understood as a consequence of the absence of Pauli exclusion effects, since the two hadrons involved do not share common valence quarks. The attractive $N-\Omega_{\rm 3c}$ also exhibit a characteristic two-component structure, similar to that observed in the $N-{\rm\bar{c}c}$ potentials~\cite{Lyu:2024ttm}, consisting of an attractive core at short distances and an attractive tail at long distances.

\begin{figure}[t]
    \centering
    \includegraphics[width=\graphsize\linewidth]{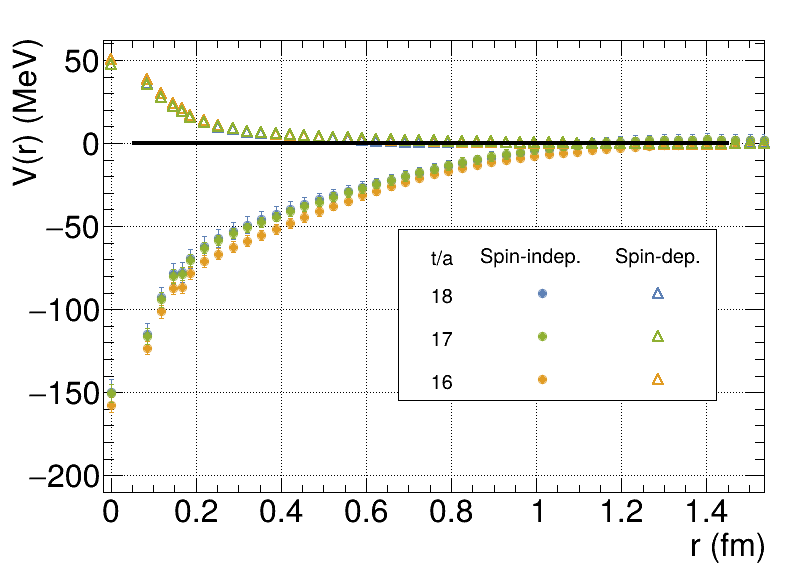}

    \caption{Spin-independent $V_0$ (full circles) and spin-dependent $V_s$ (open triangles) potentials of $N-\Omega_{\rm 3c}$, decomposed from the $^3{\rm S}_1$ and $^5{\rm S}_2$ channels, for $t/a=16$ (yellow), $17$ (green) and $18$ (blue). The figure is adapted from Ref.~\cite{Zhang:2025zaa}.}
    \label{fig:pot-spin-dep}
\end{figure}

The LO potential $V^J_{\rm LO}$ (Eq.~(\ref{eq:local-pot})) with $L=0$ can be decomposed into the spin-independent central potential $V_0$ and the spin-dependent part $V_s$ with the $^3{\rm S}_1$ and $^5{\rm S}_2$ channels.
Fig.~\ref{fig:pot-spin-dep} displays the spin-independent potential $V_0$ and the spin-dependent potential $V_s$.
The S-wave $N-\Omega_{\rm 3c}$ interaction is dominated by the spin-independent component, which provides a significant attraction, whereas the spin-dependent potential contributes only at short distances. The positive $V_s$ indicates that the $J=1$ channel is more attractive than the $J=2$ channel, reflecting the opposite signs of the spin–spin operator eigenvalues in the two total-spin states.

\subsection{Phase shifts and scattering parameters}
\label{sub:phase-shift}

To obtain physical observables in the infinite-volume limit, the lattice potentials are parametrized by a phenomenological two-range Gaussian form, 
$V_{\rm fit}(r)=\sum_{i=1}^2 A_i\exp\left(-(r/B_i)^2\right)$,
which effectively captures the characteristic two-component structure of the $N-\Omega_{\rm 3c}$ interaction. As a representative example, Fig.~\ref{fig:pot-spin-channels} shows the uncorrelated fits for both spin channels at $t/a=17$ in the range $0<r<3$~fm, while the resulting fit parameters are summarized in Table~\ref{tab:fitparameter}.

\begin{table}[t]
    \centering
\caption{Parameters of the two-range Gaussian fit $V_{\rm fit}$ for the $N-\Omega_{\rm 3c}$ potentials.}
\label{tab:fitparameter}
\begin{tabular}{crrrr}
    \multicolumn{5}{l}{$^3{\rm S}_1$ channel}\\
    \toprule
    $t/a$ & $A_1$~[MeV] & $A_2$~[MeV] & $B_1$~[fm] & $B_2$~[fm]  \\
    \midrule
        16 & $-118.9(1.6)$& $-85.7(2.6)$ & $0.142(7)$ & $0.633(33)$ \\
        17 & $-118.0(3.0)$ & $-80.0(3.7)$ & $0.135(8)$ & $0.601(37)$ \\
        18 & $-119.7(4.6)$ & $-75.0(8.5)$ & $0.141(12)$ & $0.608(56)$ \\
    \bottomrule
    \multicolumn{5}{l}{}\\
    \multicolumn{5}{l}{$^5{\rm S}_2$ channel}\\
    \toprule
    $t/a$ & $A_1$~[MeV] & $A_2$~[MeV] & $B_1$~[fm] & $B_2$~[fm]  \\
    \midrule
        16 & $-50.5(3.6)$ & $-66.5(3.0)$ & $0.110(15)$ & $0.665(6)$ \\
        17 & $-52.6(2.5)$ & $-60.4(2.5)$ & $0.110(12)$ & $0.612(50)$ \\
        18 & $-53.0(5.2)$ & $-57.8(5.2)$ & $0.113(21)$ & $0.636(67)$ \\
    \bottomrule
\end{tabular}
\end{table} 

\begin{figure}[t]
    \centering
    \includegraphics[width=\graphsize\linewidth]{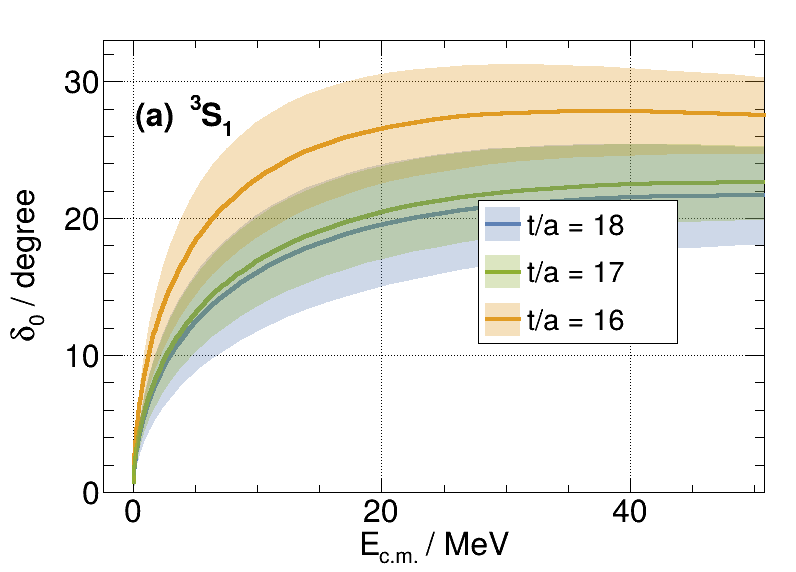}


    \includegraphics[width=\graphsize\linewidth]{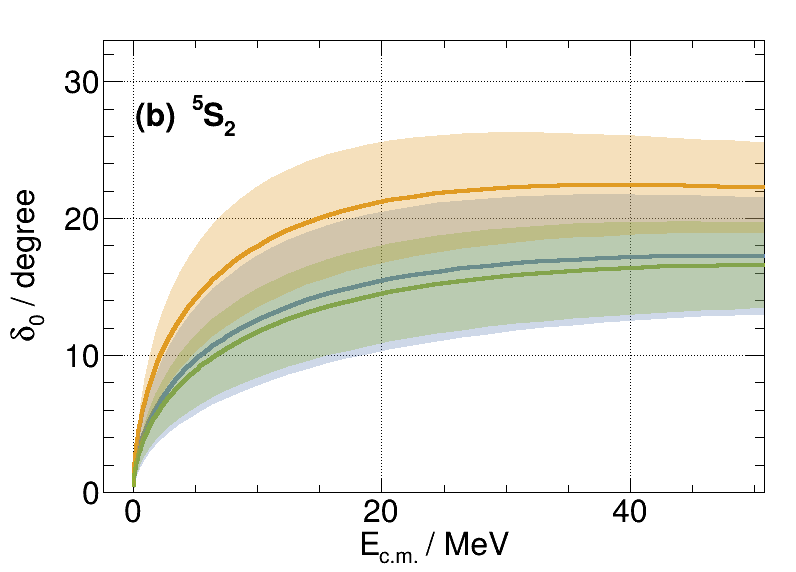}
    \caption{S-wave $N-\Omega_{\rm 3c}$ scattering phase shifts for $t/a=16$ (yellow), $17$ (green) and $18$ (blue). The solid curves represent the central values, while the shaded bands indicate the statistical uncertainties. Panel (a) shows the $^3{\rm S}_1$ channel, and panel (b) shows the $^5{\rm S}_2$ channel. The figure is adapted from Ref.~\cite{Zhang:2025zaa}.}
    \label{fig:phaseshift}
\end{figure}

Fig.~\ref{fig:phaseshift} presents the S-wave $N-\Omega_{\rm 3c}$ scattering phase shifts $\delta_0$ as functions of the center-of-mass kinetic energy $E_{\rm c.m.}=k^2/(2\mu)$ for $t/a=16,\ 17,$ and $18$. The phase shifts are obtained by solving the infinite-volume Schr\"odinger equation using the fitted potential $V_{\rm fit}(r)$. In both spin channels, $\delta_0$ approaches $0^\circ$ in the $k\to0$ limit, indicating the absence of bound states below the $N-\Omega_{\rm 3c}$ threshold. In the following, we quote the central values and statistical errors at $t/a=17$ and estimate the systematic uncertainty from the variation among $t/a=16,\ 17,$ and $18$.

\begin{table}
    \centering
\caption{The $N$-$\Omega_{\rm 3c}$ scattering length $a_0$ and effective range $r_{\text{eff}}$. Central values are taken from $t/a=17$, and
statistical uncertainties are evaluated using the jackknife method, while systematic uncertainties are estimated from the variation among $t/a=16,\ 17,$ and $18$.}
\label{tab:scatteringpara}
    \begin{tabular}{ccc} 
        \toprule
         channel&  $a_0$~[fm]& $r_{\text{eff}}$~[fm]\\ 
        \midrule
            $^3{\rm S}_1$&  $0.56(0.13)\left(^{+0.26}_{-0.03}\right)$& $1.60(0.05)\left(^{+0.04}_{-0.12}\right)$\\ 
            $^5{\rm S}_2$&  $0.38(0.12)\left(^{+0.25}_{-0.00}\right)$& $2.04(0.10)\left(^{+0.03}_{-0.22}\right)$\\
        \bottomrule
    \end{tabular}
\end{table}

Low-energy scattering parameters are determined from the effective range expansion (ERE) of the phase shifts up to next-to-leading order,
$k\cot\delta_0 = \frac{1}{a_0}+\frac{1}{2}r_{\rm eff}k^2+\mathcal{O}(k^4),$
where $a_0$ and $r_{\rm eff}$ denote the scattering length and effective range, respectively.
The extracted values are summarized in Table~\ref{tab:scatteringpara}.

\section{Discussions}
\label{discuss}

In this section, we compare the present $N-\Omega_{\rm 3c}$ results with those for the related $N-J/\psi$ and $N-\Omega_{\rm 3s}$ systems studied previously, aiming to clarify common features and differences in heavy-hadron interactions.
The $N-\Omega_{\rm 3c}$ analysis is based on the F-conf ($m_\pi\simeq137$ MeV), while the $N$–$J/\psi$ and $N-\Omega_{\rm 3s}$ results were obtained using the K-conf ($m_\pi\simeq146$ MeV)~\cite{Ishikawa:2015rho}.
As a result, the following comparison should be regarded as semi-quantitative.
A fully consistent comparison using a single gauge ensemble will be presented in a future study.

\subsection{Comparison with \texorpdfstring{$N-\Omega_{\rm{3s}}$}{N-Omegasss} potential}
\label{sub:vs-VNO3s}

\begin{figure}[t]
    \centering
    \includegraphics[width=\graphsize\linewidth]{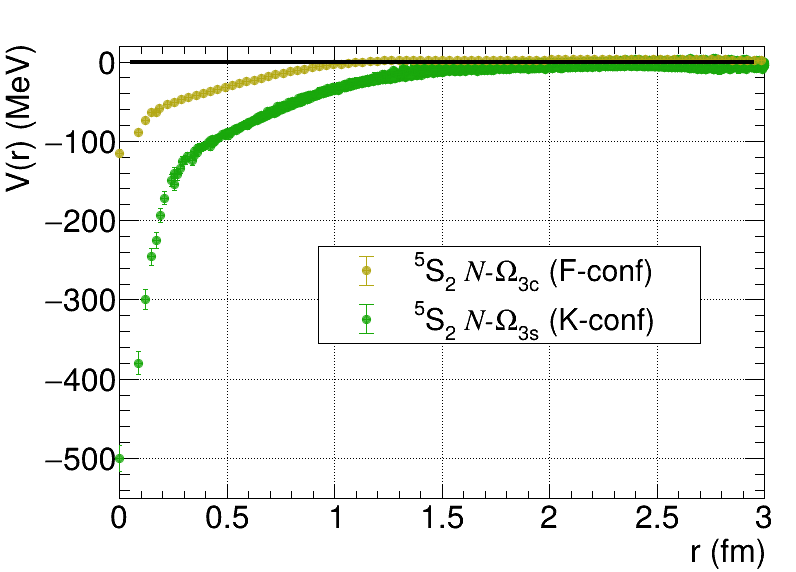}

    \caption{Comparison of the $^5{\rm S}_2$ $N-\Omega_{\rm 3c}$ potential from F-conf ($t/a=17$) with the $^5{\rm S}_2$ $N-\Omega_{\rm 3s}$ potential from K-conf ($t/a=14$)~\cite{HALQCD:2018qyu}. The $N-\Omega_{\rm 3c}$ system exhibits a weaker short-range attraction, in qualitative agreement with phenomenological quark-model expectations~\cite{Oka:1986fr}. The figure is adapted from Ref.~\cite{Zhang:2025zaa}.}
    \label{fig:vs-VNO3s}
\end{figure}

A quasi-bound state was observed in the $^5{\rm S}_2$ $N-\Omega_{\rm 3s}$ channel~\cite{HALQCD:2018qyu}, while there is no bound state in $N-\Omega_{\rm 3c}$ as discussed in Sec.~\ref{sub:phase-shift}. This reflects the overall weaker attraction in the $N-\Omega_{\rm 3c}$ system. As a qualitative comparison shown in Fig.~\ref{fig:vs-VNO3s}, the $^5{\rm S}_2$ $N-\Omega_{\rm 3c}$ potential obtained with the F-conf has a similar shape but 2-5 times less attractive to the $^5{\rm S}_2$ $N$-$\Omega_{\rm {3s}}$.

This difference can be understood in terms of two main effects.
At long distances, the interaction is dominated by meson exchange, where two-$K$ exchange in the $N-\Omega_{\rm 3s}$ system is more attractive and longer-ranged than two-$D$ exchange in the $N-\Omega_{\rm 3c}$ system due to the lighter meson mass.
At short distances, the chromo-magnetic interaction~\cite{Oka:1986fr}, which scales inversely with the constituent quark mass, is suppressed for charm quarks, leading to a weaker attraction in the $N-\Omega_{\rm 3c}$ system compared to $N-\Omega_{\rm 3s}$.

\subsection{Comparison with \texorpdfstring{$N$-$J/\psi$}{N-Jpsi} potential}
\label{sub:vs-VJpsi}

\begin{figure}[t]
    \centering
    \includegraphics[width=\graphsize\linewidth]{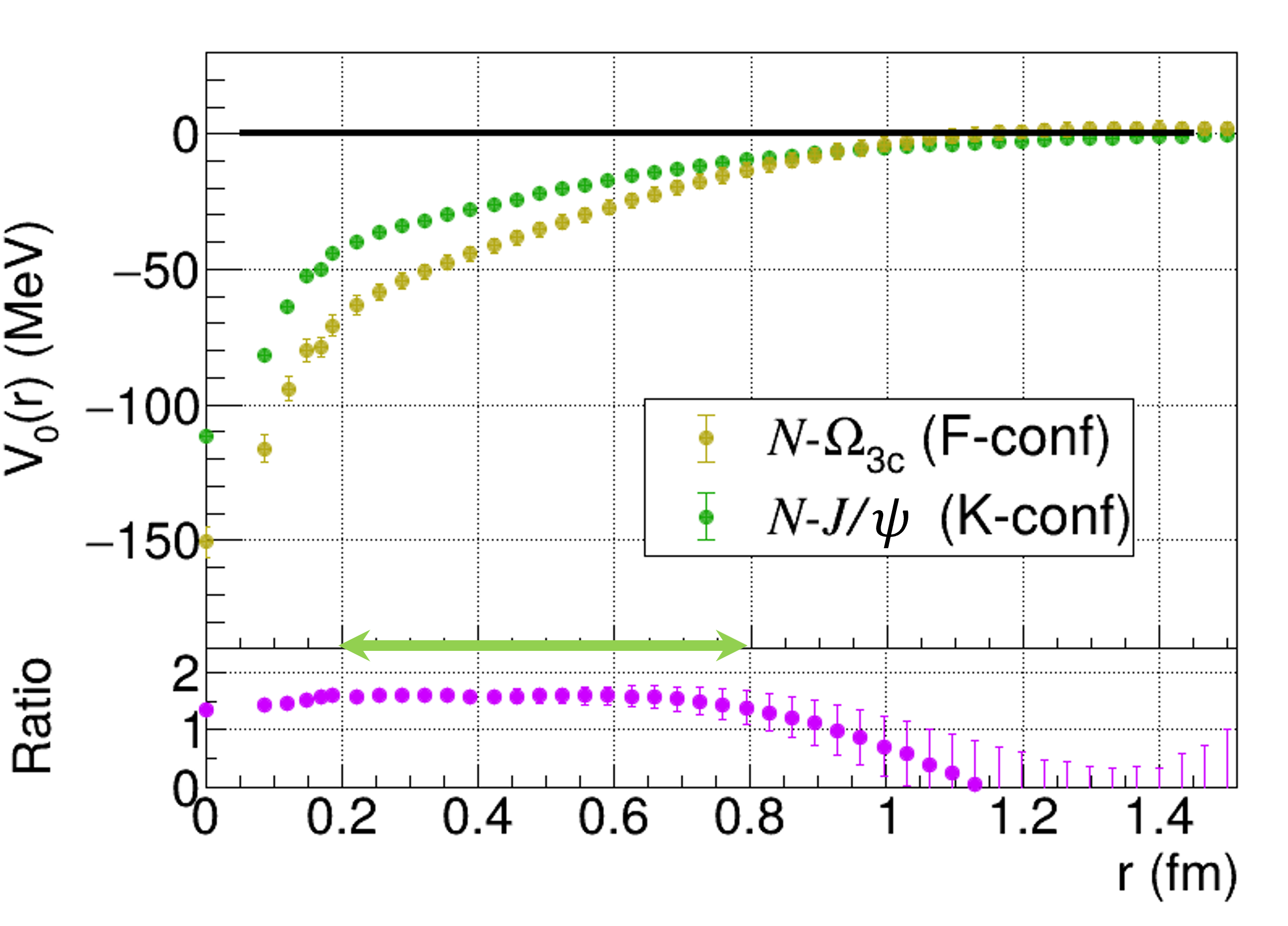}

    \caption{Comparison of the spin-independent potentials for $N-\Omega_{\rm 3c}$ from F-conf (at $t/a=17$)
    and $N-J/\psi$ from K-conf (at $t/a=13$)~\cite{Lyu:2024ttm}. The lower panels showing their ratio. The interval indicated by the green arrow exhibits a clear plateau, suggesting a same interaction mechanism in the two systems. The figure is adapted from Ref.~\cite{Zhang:2025zaa}.
    }
    \label{fig:vs-VNJpsi}
\end{figure}


The $N-\Omega_{\rm 3c}$ spin-independent potential shows a clear similarity to the $N-J/\psi$, as illustrated in Fig.~\ref{fig:vs-VNJpsi}.
This resemblance provides a qualitative indication that the long-range interactions in both systems may be governed by a common mechanism.
In particular, soft-gluon exchange between the nucleon and the heavy hadron can be effectively interpreted as a two-pion exchange interaction at long distances~\cite{Hatsuda:2025djd}, providing a natural explanation for the observed similarity.

The ratio of the two potentials, lower panel in Fig.~\ref{fig:vs-VNJpsi}, exhibits a plateau over an intermediate distance range. The plateau indicates that the relative coupling strengths of $\Omega_{\rm 3c}$ and $J/\psi$ to soft gluonic fields are approximately constant.
This observation further supports the interpretation that the $N-\Omega_{\rm 3c}$ and $N-J/\psi$ interactions share a common long-range dynamics driven by soft-gluon exchange, which can be effectively modeled by a two-pion exchange mechanism at long distances~\cite{Hatsuda:2025djd}.

\section{Summary}
\label{summary}

In this work, we studied the $N-\Omega_{\rm 3c}$ interaction using $(2+1)$-flavor lattice QCD simulations at the physical point and the time-dependent HAL QCD method~\cite{Zhang:2025zaa}.
Effective interactions in the $^3{\rm S}_1$ and $^5{\rm S}_2$ channels were extracted in terms of potentials and scattering phase shifts.

We find that the $N-\Omega_{\rm 3c}$ interaction is attractive in both spin channels.
The interaction is dominated by the spin-independent potential $V_0$, which is attractive over all distances, while the spin-dependent potential $V_s$ contributes at short distances.
The resulting phase shifts indicate that no bound state exists in the $N-\Omega_{\rm 3c}$ system in either channel.

A comparison with related systems shows that the $N-\Omega_{\rm 3c}$ potential is significantly weaker than the $N$–$\Omega_{\rm 3s}$ potential. This reduction in attraction can be attributed to the larger charm-quark mass compared to the strange-quark mass, which suppresses both meson-exchange contributions and short-range chromo-magnetic interactions.

On the other hand, the spin-independent potential $V_0$ of the $N-\Omega_{\rm 3c}$ system exhibits a striking similarity to that of the $N$–$J/\psi$ system at intermediate distances.
This observation suggests that both interactions may be governed by a common mechanism, plausibly originating from soft-gluon exchange, which effectively reduces to a two-pion exchange interaction.

This physical point lattice QCD study provides new insight into heavy-baryon interactions and offers a quantitative basis for future phenomenological and experimental investigations involving charm hadrons.

\section*{Acknowledgements}


We thank the HAL QCD Collaboration for valuable discussions.
This work was supported by the HPCI System Research Project, JSPS, RIKEN (including Fugaku-related programs), JICFuS, JST ASPIRE, the China Scholarship Council, NSFC, and regional funding agencies in China.
The lattice QCD calculations were performed on the Fugaku and HOKUSAI supercomputers at RIKEN.

\appendix

\section{Definition of baryon operators}
\label{App:operators}

The nucleon and $\Omega_{\rm 3c}$ sink operators at time $t$ are:
\begin{equation}
    \begin{aligned}
        &\left[N\right]_{\alpha}(\boldsymbol{x})=\epsilon_{abc}\left(u^{a^T}(\boldsymbol{x}){\rm C}\gamma_5d^b(\boldsymbol{x})\right)q_\alpha^c,\\
        &\left[\Omega_{\rm 3c}\right]_{\beta l}(\boldsymbol{x})=\epsilon_{abc}\left({Q^{a^T}}(\boldsymbol{x}){\rm C}\gamma_lQ^b(\boldsymbol{x})\right)Q_\beta^c,
    \end{aligned}
\end{equation}
where $q=(u,d)^T$ representing the light quark field and $Q$ being the charm quark field. $\alpha$ and $\beta$ are spinor indices, 
$l$ is a spatial index for the gamma matrix, and $a,b,c$ are color indices. ${\rm C}$ is the charge conjugation matrix.

\section{Explicit form of the spin projection operators}
\label{App:Ps}

In this appendix, we present the explicit index representation of the spin projection operators used in the construction of the $N\Omega_{\rm 3c}$ system, based on the definition of baryon operators (\ref{App:operators}). The spin projection operator $P^s$ is implemented by
\begin{equation}
\label{eq:Ps}
\begin{aligned}
    P^{(s=1)}=P_{\Omega_{\rm 3c}}^{s=3/2}\frac{1}{2}\left(\frac{3}{4}-\boldsymbol{S}_N\cdot \boldsymbol{S}_{\Omega_{\rm 3c}}\right)P_{\Omega_{\rm 3c}}^{s=3/2}\\
    P^{(s=2)}=P_{\Omega_{\rm 3c}}^{s=3/2}\frac{1}{2}\left(\frac{5}{4}+\boldsymbol{S}_N\cdot \boldsymbol{S}_{\Omega_{\rm 3c}}\right)P_{\Omega_{\rm 3c}}^{s=3/2}
\end{aligned}
\end{equation}
in which $\boldsymbol{S}_N\cdot\boldsymbol{S}_{\Omega_{\rm 3c}}$ is the spin–spin operator acting on the nucleon–$\Omega_{\rm 3c}$ system, and $P_{\Omega_{\rm 3c}}^{s=3/2}$ ensures a pure spin-$3/2$ $\Omega_{\rm 3c}$ baryon operator.
The explicit matrix representations of each components are given by:
\begin{equation}
\label{App-eq:Ps}
    \begin{aligned}
        &(\boldsymbol{S}_N\cdot \boldsymbol{S}_{\Omega_{\rm 3c}})_{\alpha\beta l;\alpha'\beta' l'}=
            \sum_i\frac{1}{4} \delta _{l l'} \sigma _{\alpha \alpha'}^i \sigma_{\beta \beta'}^i
            -\frac{1}{2} i\epsilon _{i l l'}  \sigma _{\alpha \alpha'}^i \delta _{\beta \beta'},\\
        &\left(P_{\Omega_{\rm 3c}}^{s=3/2}\right)_{\alpha\beta l;\alpha'\beta'l'}=
            \delta_{\alpha \alpha'}
            \left(\delta _{\beta  \beta'} \delta _{l l'}-\frac{1}{3} \sigma _{\beta  \rho }^l \sigma _{ \rho \beta'}^{l'}\right).
    \end{aligned}
\end{equation}
Here, $\sigma^i$ are Pauli matrices and the indexes are same with \ref{App:operators}.

\bibliographystyle{elsarticle-num}
\bibliography{02_2}






\end{document}